\documentclass[12pt]{article}
\usepackage{graphicx}
\usepackage{latexsym}
\def\tr{{\rm tr}}
\def\diag{{\rm diag}\,}
\def\Tr{{\rm Tr}}

\def\C{{\bf C}}
\def\Z{{\bf Z}}
\def\R{{\bf R}}

\def\CD{{\cal D}}
\def\CE{{\cal E}}

\def\CM{{\cal M}}

\def\CL{{\cal L}}
\def\CN{{\cal N}}
\def\CO{{\cal O}}
\def\CP{{\cal P}}
\def\CQ{{\cal Q}}

\def \slash#1{\centeron{$#1$}{$/$}}
\def\centeron#1#2{{\setbox0=\hbox{#1}\setbox1=\hbox{#2}\ifdim
   \wd1>\wd0\kern.48\wd1\kern-.48\wd0\fi
   \copy0\kern-.48\wd0\kern-.48\wd1\copy1\ifdim\wd0>\wd1
   \kern.48\wd0\kern-.48\wd1\fi}}

\newcommand{\beq}{\begin{equation}}
\newcommand{\eeq}{\end{equation}}
\newcommand{\bea}{\begin{eqnarray}}
\newcommand{\eea}{\end{eqnarray}}
\newcommand{\ba}{\begin{array}}
\newcommand{\ea}{\end{array}}

\newcommand{\p}{\partial}
\newcommand{\nn}{\nonumber}

\newcommand{\half}{\frac{1}{2}}

\begin{document}
\vfill
\begin{flushright}
{\tt\normalsize KIAS-P08083}
\end{flushright}
\hskip3cm

\centerline{\LARGE \bf  Membrane Dynamics in Three dimensional }
\vskip0.5cm \centerline{\LARGE \bf ${\cal N}=6$ Supersymmetric Chern-Simons Theory}

\vskip2cm

\centerline{\Large Jong-Hyun Baek\footnote{jbaek@yonsei.ac.kr}, Seungjoon Hyun\footnote{ hyun@phya.yonsei.ac.kr}, Wooje Jang\footnote{wooje@yonsei.ac.kr} and
Sang-Heon Yi\footnote{shyi@phya.yonsei.ac.kr}}

\hskip2cm

\begin{center}
{\it Department of Physics, College of Science, Yonsei University,
Seoul 120-749, Korea \\
${}^2$Korea Institute for Advanced Study, Seoul 130-722, Korea}
\end{center}

\hskip2cm


\vskip2cm

\centerline{\bf Abstract}
We study the membrane scattering in the three-dimensional ${\cal N}=6$ supersymmetric Chern-Simons theory recently constructed by Aharony, Bergman, Jafferis and Maldacena and conjectured to be dual to M-theory on $AdS_4\times S^7/\Z_k$. We compute the one-loop effective action up to the $v^4$ terms in the derivative expansion and find exact agreement with the results from the supergravity computations. In particular, our results imply that the $v^2$ term is not renormalized and tree-level exact.  
\thispagestyle{empty}
\renewcommand{\thefootnote}{\arabic{footnote}}
\setcounter{footnote}{0}
\newpage

\section{Introduction}
Recently there have been much interests in the worldvolume theory of multi M2 branes which is dual to M-theory on $AdS_4\times S^7$ spacetime, initiated by the pioneering works of Bagger and Lambert~\cite{Bagger:2007vi}\cite{Bagger:2007jr} and Gustavsson~\cite{Gustavsson:2007vu} (BLG) based on the, so called, 3-algebra. Especially, the three-dimensional ${\cal N}=6$ superconformal $U(N)\times U(N)$ Chern-Simons gauge theory with level $(k, -k)$ constructed by Aharony, Bergman, Jafferis and Maldacena (ABJM)~\cite{Aharony:2008ug} is conjectured to be dual to M-theory  on $AdS_4\times S^7/ \Z_k $. Some evidences supporting the conjecture include the fact that the  classical moduli space is given by $C^4/\Z_k$ and the model has $\CN=6$ superconformal symmetry. 
Since  the model is described by a single parameter,  the quantized level $k$, one may expect the conformal symmetry persists at the quantum level.  Further works have been made on  the superconformal index~\cite{Bhattacharya:2008bja},  integrability structures~\cite{Minahan:2008hf}, Wilson loops~\cite{Drukker:2008zx}, non-perturbative monopole instanton~\cite{Hosomichi:2008ip} and relation to the BLG model~\cite{Honma:2008ef}.

One nontrivial test for the ABJM model as the dual field theory of M theory on $AdS_4\times S^7/\Z_k$  is the study of the membrane scattering amplitude. In the dual gravity description,  it would be given by the effective action of the probe M2 brane in the  $AdS_4\times S^7/\Z_k$ background due to the large number of source M2 branes. See \cite{Verlinde:2008di} for the membrane scattering in the context of the BLG model.

For a slowly moving probe membrane, with constant velocity $v^I=\frac{d X^I}{d X^0}$, the effective action in the static gauge can be expressed as the derivative expansions in transverse coordinates:
\[
\Gamma=l_p^{-3}\int d^3\xi \sum_{n=0}^\infty \Big(a_n v^{2n+2} +\cdots\Big)\,,
\]
where $\cdots$ denotes the superpartners of $v^{2n}$ terms. The coefficients $a_n$ depend on the distance $r$ between the source and probe branes. Because of dimensional reasons, their leading terms are expected to be of the form 
\[
a_n \sim \Big(\frac{l_p}{r}\Big)^{6n}\,.
\] 
In the dual field theory description of M theory on $AdS_4\times S^7$,  the computation was done in~\cite{Aharony:1996bh}-\cite{Hyun:1999hf} using the worldvolume theory of multi D2 branes, which is the three-dimensional ${\cal N}=8$ super Yang-Mills theory, and taking the strong coupling limit or decompactification limit of M-circle. It was found that the $v^2$ term and its superpartners are tree-level exact and the  $v^4$ term  is given by the sum of one-loop correction and the infinite sum of monopole instanton corrections. They match exactly with the results from the supergravity~\cite{Hyun:1998qf}. 
The exactness of these results is due to the ${\cal N}=8$ supersymmetry~\cite{Dine:1997nq}\cite{Paban:1998qy}.

One may expect similar behavior in the ABJM model. The model has four complex bifundamental scalar fields $Y^A$ which has mass dimension $\half$. The vacuum expectation values $b^A\equiv \langle Y^A\rangle$ span the vacuum moduli space, which corresponds to the transverse space of the membranes. Since the only dimensionful parameter at a generic point of moduli space is the vev, $b$,  the superconformal invariance would demand the effective action to have the form
\[
\Gamma=\int d^3\xi \sum_{n=0}^\infty\Big[ c_n v^2\Big(\frac{v^2}{b^6}\Big)^{n} +\cdots\Big]\,,
\] 
where $c_n$ is a dimensionless constant~\cite{Berenstein:2008dc}. It has been known that the $v^2$ term is one-loop exact with $\CN=4$ supersymmetry~\cite{Seiberg:1996bs}, and tree-level exact with  $\CN=8$ supersymmetry~\cite{Paban:1998ea}. This means that we can determine the $v^2$ term exactly by studying one-loop corrections and find out its (non)renormalization.
It is also known that the $v^4$ term is one-loop exact, apart from the possible nonperturbative instanton corrections in the   $\CN=8$ supersymmetric field theories~\cite{Dine:1997nq}. Therefore, by studying one-loop corrections, we can determine the $v^4$ term exactly, at least for level $k=1, 2$ where the supersymmetry is believed to be enhanced to $\CN=8$. 

In this paper, we compute the one loop corrections to the effective action. We find there is  no one-loop correction in the $v^2$ term and thus it is tree-level exact. We also obtain the $v^4$ term in an expected form at one-loop. This should be an exact result  if we interpolate level $k$ to 1 or 2. 
It would be very nice to see if it is true for general $k$ by using supersymmetry arguments. In any case, these results exactly agree with those from supergravity computations, which supports the correspondence between the ABJM model and M theory on $AdS_4\times S^7/\Z_k$.

In section 2, we summarize the relevant results from the supergravity. In section 3, we review briefly the ABJM model and then present the basic set-up for our computations including the background configurations and the gauge fixing.  In section 4, we present our one-loop calculations. Since the computations are somewhat involved, we mainly focus on the results while the details are deferred  to the appendices.  In section 5, we give some concluding remarks. In appendix A, we present the gauge fixing for an alternative choice of background configuration. In appendix B, we give details on the one-loop calculations. In appendix C, we present the relevant integrations.

\section{The results from supergravity}

The eleven dimensional metric describing $N$ M2 branes is given by 
\begin{equation}
\label{m2metric}
ds^2_{11} = h^{-2/3} (-dt^2+dx_9^2+dx_{10}^2)
       +h^{1/3} (dx_1^2 + \cdots +dx_{8}^2) , 
\end{equation}
where $h$ is the transverse eight dimensional harmonic function of $r= (  x_1^2 + \cdots + x_{8}^2 )^{1/2}$,
\begin{equation}
h(r) = 1+  \frac{32\pi^2 Nl_p^6}{r^6} . 
\label{m2harmonic}
\end{equation}
The near-horizon limit of (\ref{m2metric}) is given by  the  $AdS_4\times S^7$ geometry in which the harmonic function becomes
\begin{equation}
h(r)=  \frac{32\pi^2 Nl_p^6}{r^6} . 
\label{m2harmonic}
\end{equation} 
The geometry is maximally supersymmetric with 32 Killing spinors and has  the isometry $SO(2,3)\times SO(8)$. 
This is the limit where the worldvolume theory of $N$ M2 branes is expected to become the three dimensional ${\cal N}=8$ superconformal field theory. 

In addition, one can consider the $\Z_k$ orbifolding to the transverse space, $\R^8/\Z_k=\C^4/\Z_k$, with $(x^A + i x^{A+4})\sim e^{  i\frac{2\pi}{k}}( x^A + i x^{A+4})$, $A=1,\cdots, 4$.
The effect of $\Z_k$ orbifolding corresponds to the replacement $N\rightarrow N'=kN$ in the harmonic function $h(r)$~\cite{Aharony:2008ug}. In the near horizon limit, the geometry admits 24 Killing spinors and has the isometry  $SO(2,3)\times SO(6)$ for $k\geq 3$. The dual field theory is expected to be an $\CN=6$ superconformal theory. 

The Nambu-Goto action for a probe membrane is given by
\begin{equation}
S_2=T_2\Big( -\int d^3\xi \sqrt{ - \det h_{\mu\nu}} + \int H \Big)~,
\end{equation}
where $T_2=1/(4\pi^{2} l_p^{3})$ is the membrane tension and $h_{\mu\nu}$ is the induced metric on the worldvolume. If one uses the static gauge for worldvolume diffeomorphism, the induced metric becomes 
\begin{equation}
h_{\mu\nu}=g_{\mu\nu}+\partial_\mu X^I \partial_\nu X^J g_{IJ}~,
\end{equation}
where $X^I$ ($I= 1, \cdots, 8$) are transverse coordinates. We consider the configurations
in which the probe membrane is parallel to the source membranes and is scattered with a constant velocity.  Furthermore we restrict ourselves to the case that $X^I$ depends only on time.

After plugging the $AdS_4\times S^7/\Z_k$ metric into the probe action and expanding it in terms of the velocity, $v^I  = \dot{X}^I$, we obtain the effective action of the form
\begin{equation}
S_2=\int d^3\xi \Big( \frac{1}{2} T_2v^2 -V_2 +{\cal O}(v^6)\Big)\,,
\end{equation}
where $V_2$, the interaction potential of $v^4$ order, is given by
\begin{eqnarray}
V_2 = -\frac{1}{8} T_2 h(r) (v^2)^2 = - kN l_p^{3}\frac{ (v^2)^2}{r^6}~. \label{sugrapot}
\end{eqnarray}

To compare with the results from the ABJM model, we introduce complex coordinates in the transverse space as
\begin{eqnarray}
z^A=\frac{1}{2 \sqrt{\pi k l_p^3}}\Big( X^A + i X^{A+4}\Big), \qquad A=1, \cdots, 4~.
\end{eqnarray}
where the orbifolding $\Z_k$ acts as 
$ z^A \rightarrow  e^{i\frac{2\pi}{k}}z^A
$.
In these coordinates, the effective action becomes
\begin{equation}
S_2=\int d^3\xi \Big( \frac{k}{2\pi } |v|^2+\frac{N}{4\pi}\frac{|v|^4}{|z|^6} +{\cal O}(|v|^6)\Big)~,
\end{equation}
with $|z|^2\equiv z^A z_{A}^ *$ and $|v|^2\equiv \dot{z}^A \dot{z}_{A}^ *$.

\section{Three-dimensional ${\cal N}=6$ superconformal Chern-Simons theory}

The ABJM model has  $OSp(6|4)$ superconformal symmetry for generic $k$. The classical vacuum moduli space is given by $\C^4/\Z_k$. These indicates that the model could be  the worldvolume theory of $N$ M2-branes, which is dual to $M$ theory on $AdS_4\times S^7/\Z_k$.  It is curious to note that this conjectured duality implies that the model has enhanced supersymmetries to $\CN=8$ for $k=1,2$. 
In section 3.1, we give some review on the model, establishing our notation. In section 3.2, we give the basic set-up of the problem, which corresponds to the membrane scattering in the dual gravity descriptions. In section 3.3, we describe the appropriate gauge fixing for the given configuration. In section 3.4, we present the tree level quadratic Lagrangian for gauge/scalar, fermion and ghost fields, respectively.
\subsection{The ABJM Model}
The model contains scalars, fermions and gauge fields. As being a conformal field theory,  
the $U(N)\times U(N)$ gauge fields $A_L$, $A_R$ have Chern-Simons action with level $k$ and $-k$, respectively. 
The matter fields 
consist of four complex scalar fields $Y^A$ and spinor fields $\Psi_A$, which transform  as  ${\bf 4}$ and $\bar{\bf 4}$ under $SU(4)$ R-symmetry of $\CN=6$ supersymmetry. They have bifundamental representations under the gauge groups. 

Our conventions for spinors and their contractions are as follows. The three dimensional worldvolume flat metric and totally antisymmetric $\epsilon$-tensor are taken as $\eta^{\mu\nu}=\diag(-,+,+)$ and $\epsilon^{012} =1$. The three dimensional Dirac $\gamma$-matices satisfy $ \gamma^{\mu}\gamma^{\nu} = \eta^{\mu\nu} + \epsilon^{\mu\nu\rho}\gamma_{\rho}\,.$
An explicit realization 
may be given by
$ \gamma^{\mu~~ \beta}_{~~\alpha} = (i\sigma^2, \sigma^3,-\sigma^1)\,. $
Indices of three dimensional spinors are  raised or lowered by antisymmetric $\epsilon$-matrices, $\epsilon^{12} = \epsilon_{21} = 1$. 
%
%
We always contract spinor indices from northwest to southeast: 
%
\[ \psi \chi \equiv 
\psi^{\alpha}\chi_{\alpha} = \epsilon^{\alpha\beta}\psi_{\beta}\chi_{\alpha} = \chi^{\alpha}\psi_{\alpha}=\chi\psi\,.\] 
Similarly, fermion bilinears with $\gamma$-matrices are defined as 
\[ \psi\gamma^{\mu}\chi \equiv \psi^{\alpha}\gamma^{\mu~~ \beta}_{~~ \alpha}\chi_{\beta} = -\chi^{\alpha}\gamma^{\mu~~ \beta}_{~~\alpha}\psi_{\beta}=-  \chi\gamma^{\mu}\psi\,.  \] 
The hermitian conjugate is defined as
$ (\psi^{\dagger}\chi)^{\dagger} 
= \chi^{\dagger}_{\alpha}\psi^{\alpha} = - \chi^{\dagger}\psi\,.$
Note that $\gamma^{\mu}_{\,\alpha\beta} \equiv \epsilon_{\beta\rho}\gamma^{\mu~~ \rho}_{~~\alpha}=(-{\bf 1},\sigma^1,\sigma^3)$ are real and symmetric.

In order to do one loop computation, we choose the normalization of  the scalar and fermion fields so that the classical action has an overall factor of the coupling constant, $\frac{k}{2\pi}$.  In these conventions, our starting ABJM action is given by %
\bea S &=& \frac{k}{2\pi}\int d^3x\bigg[\half \epsilon^{\mu\nu\rho}\Tr \bigg\{  A_{L\, \mu}\p_{\nu}A_{L\, \rho} + \frac{2i}{3}A_{L\, \mu}A_{L\, \nu}A_{L\, \rho} - A_{R\, \mu}\p_{\nu}A_{R\, \rho}-\frac{2i}{3}A_{R\, \mu}A_{R\, \nu}A_{R\, \rho}\bigg\}   \nn \\
&&~~~~~  +\Tr\bigg\{ -(D_{\mu}Y^{\dagger}_{ A})(D^{\mu}Y^{A})+i\Psi^{\dagger\, A}\gamma^{\mu}D_{\mu}\Psi_{A}\bigg\}  -V_b-V_f\bigg]\,, \eea
where 
\bea 
V_b &=& -\frac{1}{3}\Tr\bigg[Y^{\dagger}_{A}Y^{A}Y^{\dagger}_{B}Y^{B}Y^{\dagger}_CY^C + Y^{A}Y^{\dagger}_{A}Y^{B}Y^{\dagger}_{B}Y^CY^{\dagger}_C \nn\\
&& ~~~~~~+ 4Y^{A}Y^{\dagger}_BY^{C}Y^{\dagger}_{A}Y^{B}Y^{\dagger}_C - 6Y^{A}Y^{\dagger}_{B}Y^{B}Y^{\dagger}_AY^{C}Y^{\dagger}_C \bigg]\,,\\
&& \nn \\  
V_f &=& i\Tr\bigg[Y^{\dagger}_{A}Y^{A}\Psi^{\dagger\, B}\Psi_B - Y^{A}Y^{\dagger}_A\Psi_B\Psi^{\dagger\, B}-2Y^{\dagger}_AY^{B}\Psi^{\dagger\, A} \Psi_B + 2 Y^AY^{\dagger}_B\Psi_A\Psi^{\dagger\, B} \nn \\
&& ~~~~~~-\epsilon^{ABCD}Y^{\dagger}_{A}\Psi_BY^{\dagger}_C\Psi_D   +\epsilon_{ABCD}Y^{A}\Psi^{\dagger\, B}Y^C\Psi^{\dagger\, D}\bigg]\,.  \nn 
\eea
The covariant derivatives are defined by
\bea D_{\mu}Y^{A} &=& \p_{\mu}Y^{A} + iA_{L\, \mu}Y^{A} -iY^{A}A_{R\, \mu}\,, \nn \\
  D_{\mu}\Psi_{A} &=& \p_{\mu}\Psi_{A} + iA_{L\, \mu}\Psi_{A} -i\Psi_{A}A_{R\, \mu}\,. \nn \eea

The model does not have any dimensionful parameter and thus the classical action is conformally invariant. The gauge fields have mass dimension 1, while the scalar and fermion fields have  mass dimension $\half$ and 1, respectively.  The only dimensionless parameter is the Chern-Simons level $k$,  which plays the role of the coupling constant of the model. As it is quantized, the model is expected to be conformally invariant even at the quantum level.
It also has $\CN=6$ supersymmetry. 
Supersymmetry transformations for the ABJM model in our conventions are given by~\cite{Gaiotto:2008cg}
\bea 
\delta\, Y^{A} &=& i\CE^{AB}\Psi_B\,,  \nn \\  
\delta\, Y^{\dagger}_{A} &=& i\Psi^{\dagger\, B}\CE_{AB}\,, \nn\\   
\delta\, \Psi_{A} &=& \gamma^{\mu}\CE_{AB}D_{\mu}Y^{B} - \CE_{AB}(Y^{C}Y^{\dagger}_{C}Y^{B} - Y^{B}Y^{\dagger}_{C}Y^{C}) + 2 \CE_{CD}Y^{C}Y^{\dagger}_{A}Y^{D}\,, \nn \\
\delta\,  \Psi^{\dagger\, A} &=& -\CE^{AB}\gamma^{\mu}D_{\mu}Y^{\dagger}_{B} + \CE^{AB}(Y^{\dagger}_{C}Y^{C}Y^{\dagger}_{B} - Y^{\dagger}_{B}Y^{C}Y^{\dagger}_{C}) - 2 \CE^{CD}Y^{\dagger}_{C}Y^{A}Y^{\dagger}_{D}\,, \nn \\
\delta\, A_{L
, \mu} &=& \CE^{AB}\gamma_{\mu}\Psi_{A}Y^{\dagger}_{B} + Y^{B}\Psi^{\dagger\, A}\gamma_{\mu}\CE_{AB}\,,  \\
\delta\, A_{R\, \mu} &=&  Y^{\dagger}_{B}\CE^{AB}\gamma_{\mu}\Psi_{ A}+\Psi^{\dagger\, A}\gamma_{\mu}\CE_{AB}Y^{B}\,, \nn 
\eea
where $\CE_{AB}$ and $\CE^{AB}$ are supersymmetry variation parameters and should be related as  
\[ \CE_{AB} = -\half \epsilon_{ABCD}\CE^{CD}\,, \qquad  \CE^{AB} = -\half \epsilon^{ABCD}\CE_{CD}\,, \qquad \qquad (\CE^{\alpha}_{AB})^{\dagger} = \CE^{\alpha\, AB}\,. \]
It has been argued that the supersymmetry is enhanced to $\CN=8$ when the level $k$ is 1 or 2, though this is not obvious from the field theory Lagrangian.

\subsection{Set-up}

The appropriate background configurations for a probe membrane scattered by $N$ source membranes in the dual gravity description correspond to  $$U(N+1)\times U(N+1)\rightarrow U(N)\times U(N)\times U(1)\times U(1)~.$$
For our purpose, it is enough to consider the case $N=1$.
Let us denote the vacuum expectation values and the quantum fluctuations of scalar fields as $\bar{Y}$ and $\delta Y$, respectively,   
\[ Y^{A} = \bar{Y}^{A} +   \delta Y^A\,.  \]
We make the simplest choice for the vev's as
\beq  \bar{Y}^{A} = \left( \ba{cc} 0 & 0  \\  0 & b^A \ea\right)\,, \qquad \bar{Y}^{\dagger}_{A} = \left( \ba{cc} 0 & 0  \\  0 & b^{\dagger}_A \ea\right)\,. 
\label{vev}
\eeq
For a different choice of vacua, in which instanton corrections may play some role~\cite{Hosomichi:2008ip}, see appendix A. 

To compare with the dual gravity descriptions, we restrict ourselves to the case :
\[ b^A = b_0^A + v^A t~, \]
where $b_0^A$ and $v^A$ are constants.  One may note that $b_0$ corresponds to the impact parameter in the membrane scattering and thus satisfies $b_0\cdot v^\dagger\equiv b_0^A v_A^\dagger=0$.

For a generic background $b^A$, the off-diagonal components acquire a mass of order $|b|^2$ and thus, in the low energy, they can be treated as quantum fluctuations and integrated out.
These fluctuations  are denoted as
\bea  A_{L\, \mu} &=&  \left( \ba{cc} 0 & a_{\mu}  \\  a^{\dagger}_{\mu} & 0 \ea\right)\,,  \qquad~  A_{R\, \mu}  = \left( \ba{cc} 0 & \hat{a}_{\mu}  \\  \hat{a}^{\dagger}_{\mu} & 0 \ea\right)\,,  \nn \\
   \delta Y^A &=&     \left( \ba{cc} 0 & y^A  \\  \tilde{y}^{A} & 0 \ea\right)  \,,  \qquad
 \Psi_{A} =   \left( \ba{cc} 0 & \psi_A  \\  \tilde{\psi}_{A} & 0 \ea\right)\,. \eea

We will integrate out these massive fluctuations and obtain the effective action of diagonal fields.
The resultant effective action will have $U(1)^2\times U(1)^2\times S_2$ gauge symmetry, with the permutation symmetry $S_2$ over the diagonal elements. Those abelian gauge fields can also be integrated out to give the effective action of $b^A$ and their superpartners on the moduli space $(C^4/\Z_k)^N/S_N$~\cite{Aharony:2008ug}.

\subsection{Gauge fixing}

A  convenient gauge fixing in gauge theories with matter fields in the broken phase is the, so-called, $R_\xi$ gauge as the resultant Lagrangian does not have a mixing term between the gauge and scalar fields. The gauge fixing in supersymmetric gauge theories could be even more subtle since the gauge fixing term has to preserve the supersymmetry. One way is to use the supersymmetric  $R_\xi$ gauge 
in superfield formalism~\cite{Ovrut:1981wa}. Since we are using component fields, we use the  $R_\xi$ gauge which may be supplemented with supersymmetric completion.\footnote{One may use an $\CN =2$ superfield formalism~\cite{Benna:2008zy} and use supersymmetric $R_\xi$ gauge.}
Henceforth our gauge fixing functions are given by\footnote{
$U(N)$ indices leftover are contracted and summed. For example,  $\delta Y^{\dagger}T^a\bar{Y}= \delta Y^{\dagger \hat{i} i}T^a_{ij}\bar{Y}^{j\hat{i}}$.}
\bea f_{L}^a &=& -\frac{1}{\sqrt{\xi_L}}\Big(\p_{\mu}A^{a \mu}_L +i\xi_L \delta Y^{\dagger}T^a\bar{Y} -i\xi_L \bar{Y}^{\dagger}T^a\delta Y \Big)\,,\nn \\
 f_{R}^a &=&- \frac{1}{\sqrt{\xi_R}}\Big(\p_{\mu}A^{a \mu}_R +i\xi_R \delta{Y}T^a\bar{Y}^{\dagger}  -i\xi_R \bar{Y} T^a\delta Y^{\dagger}\Big) \,, \eea
where $\xi_L$, $\xi_R$ are  arbitrary parameters with mass dimension one in {\it three dimensions}.  
The  gauge fixing  Lagrangian is given by
\beq \CL_{GF} = -\frac{1}{2} f_L^a  f_L^a - \frac{1}{2} f_R^a f_R^a\,. \eeq
One may note that in this gauge choice we do not need to introduce the Nielsen-Kallosh ghost.

This gauge fixing function is supersymmetric for the configuration (\ref{vev}) with $v=0$ if we take 
$ \xi_L =\xi_R = m_0\equiv |b_0|^2$. This is the case since we have chosen the bosonic background, i.e. $\langle \Psi\rangle =0$. The most natural choice  for the time dependent background would be to replace $b_0^A$ by $b^A$, wherever applicable. It seems also natural in view of  superfield formalism, where $b^A$ might be promoted to a superfield.
Henceforth we use the above gauge fixing function with
\[ \xi_L =\xi_R = m\equiv |b|^2 ~.\]
 If we demand the supersymmetry completion among background fields, then, after turning on $v$, we should also  include nonvanishing $\langle \Psi\rangle$ as a superpartner. 
In this case, in order to have manifest supersymmetry among background fields, the gauge fixng function is needed to have additional  terms which are bilinears in fermions,   like $\Psi^\dagger\Psi$. These will give rise  to terms in the effective action, which depend on  $\langle \Psi\rangle$ and thus give the supersymmetric completion. 
Since we are only interested in the purely bosonic terms in the effective action, we just  use the above gauge function while neglecting those terms involving fermionic background fields. Those terms in the effective action could be obtained by the supersymmetric completion of the purely bosonic terms using the supersymmetry transformation rules for background fields.

\subsection{Quadratic action for quantum fluctuations}

After this choice of gauge fixing terms, one obtains
the quadratic  Lagrangian of the bosonic  fields, $(a_{\mu}~ y^A)$ and $(\hat{a}_{\mu}~ \tilde{y}^A)$ as  
\[\CL_{b,\, quad} = -(a^{\dagger}_{\mu}~~ y^{\dagger}_{A}) \CD^{(\mu A) }_{~~~~\, (\nu\,B)} {a^{\nu}\choose y^B} -(\hat{a}^{\mu\dagger }~~ \tilde{y}^{ A}) \hat{\CD}_{(\mu A) }^{~~~~\, (\nu\,B)} ~ {\hat{a}_{\nu}\choose \tilde{y}^\dagger_B} \]
where
\bea \CD^{(\mu A) }_{~~~~\, (\nu\,B)} &=& \left(\ba{cc}  -\p^{\mu}\frac{1}{m}\p_{\nu} +  \epsilon^{\mu}_{~\nu\rho}\p^{\rho} + m\eta^{\mu}_{\nu} & 2i\p^{\mu}b^\dagger_{B} \\  -2i \p_{\nu}b^{A}& (-\Box   + m^2)\delta^{A}_{B} \ea\right)\,, 
\eea
and
\bea \hat{\CD}_{(\mu A) }^{~~~~\, (\nu\,B)}&=& \left(\ba{cc} -\p_{\mu}\frac{1}{m}\p^{\nu} -  \epsilon_{\mu}^{~\nu\rho}\p_{\rho} + m\eta_{\mu}^{\nu} &  2i\p_{\mu}b^{B} \\ -2i\p^{\nu}b^\dagger_{A}  & (-\Box +  m^2)  \delta_{A}^{~B}\ea\right)\,.  \eea

Ghost fields are introduced in the standard way as 
\bea C_L =  \left( \ba{cc} 0 &  c_L\\  \tilde{c}_L & 0 \ea\right)\,, \qquad C_R =  \left( \ba{cc} 0 &  c_R \\   \tilde{c}_R & 0 \ea\right)\,. \eea
The quadratic Lagrangian for  ghost fields becomes
\beq \CL_{g, quad} = -c^{\dagger}_{L}\CD^{g}_{L}c_{L} - \tilde{c}^{\dagger}_{L} \tilde{\CD}^{g}_{L}\tilde{c}_{L}  -c^{\dagger}_{R}\CD^{g}_{R}c_{R} - \tilde{c}^{\dagger}_{R} \tilde{\CD}^{g}_{R}\tilde{c}_{R}\,, \eeq
where
\beq \CD^{g}_{L} = \tilde{\CD}^{g}_{L} = \CD^{g}_{R} = \tilde{\CD}^{g}_{R} =\frac{1}{\sqrt{m}}(-\Box + m^2)\,. \eeq

The quadratic Lagrangian of fermionic fields becomes
\[ \CL_{f, quad} = -\psi^{\dagger\, A}\CD^{~~ B}_{A}\psi_B -  \tilde{\psi}^{\dagger\, A}\tilde{\CD}^{~~ B}_{A}\tilde{\psi}_B\,,\]
where
\bea 
(\CD^{~~ B}_{A})_{\alpha}^{~ \beta} &=& -i\delta^{~ B}_{A}\slash{\p}_{\alpha}^{~\beta} + i(m\delta^{~B}_{A}-2b^\dagger_{A}b^{B})\delta_{\alpha}^{~\beta} \,, \nn \\
(\tilde{\CD}^{~~ B}_{A})_{\alpha}^{~\beta}&=& -i\delta^{~ B}_{A}\slash{\p}_{\alpha}^{~\beta} - i(m\delta^{~B}_{A}-2b^\dagger_{A}b^{B})\delta_{\alpha}^{~\beta} \,,\eea
and $\alpha,\beta$ denote spinor indices.

In order to compute one-loop corrections, we need to rescale the gauge and ghost fields such that their kinetic terms are in standard forms. Henceforth, we perform the following time-dependent rescaling,
\[ \frac{1}{\sqrt{m}} a_{\mu} \longrightarrow a_\mu\,, \qquad \frac{1}{m^{1/4}}c_L\longrightarrow  c_L\,, \qquad \frac{1}{m^{1/4}}\tilde{c}_L\longrightarrow  \tilde{c}_L\,, \]
in which the gauge fields and ghost fields have the same mass dimension as the scalar fields. There are no extra contributions from the path integral measure due to this rescaling since we adopt dimensional regularization.

In addition to the rescaling, we perform the Wick rotation to the Euclidean space. 
Time-independent part of the scalar vev's is denoted as 
$ m_0 \equiv |b_0|^2\,. $
Then the mass parameter $m$ is given by
\[ m = m_0 + |v|^2\tau^2 \,. \]
One may regard the velocity $v$ has  small magnitude and treat it as a perturbation parameter.

The quadratic  operators, relevant to the one-loop computation, become of the form 
$$Q=Q_0+Q_1$$
where $Q_0=Q_0(m_0)$ and $Q_1=Q_1(v)$. They consist of three parts, each from gauge/scalar fields, fermionic fields and ghost fields. We present those operators below, keeping terms only up to quartic order in $v$. 

\underline{\it Gauge/scalar fields} 
\bea
Q^b_0  &=& \left(\ba{cc} -\p^{\mu}\p_{\nu} + im_0\epsilon^{\mu}_{~\nu\rho}\p^{\rho} + m^2_0 \delta^{\mu}_{\, \nu}  & 0 \\ 0 & (-\Box + m^2_0)\delta^{A}_{B}  \ea\right)\,,
\nn \\ && \nn \\
\qquad Q^b_1 &=& \left(\ba{cc}C^\mu_{\,\nu} & D^{\mu\dagger}_B \\ D_\nu^A & E^A_B\ea\right) \,,  \eea
where
\begin{eqnarray}
C^\mu_{\,\nu} &=&  i|v|^2\tau^2\epsilon^{\mu}_{~ \nu\rho}\p^{\rho}+ M(\tau)\, \delta^{\mu}_{\,\nu} + \Big(-\frac{|v|^2}{ m_0}+ 4\frac{ |v|^4\tau^2}{m_0^2}\Big) \delta^{\mu}_\tau \delta_{\nu}^\tau-\frac{|v|^2 \tau}{m_0}\Big(\p^\mu\delta^\tau_\nu-\p_\nu\delta_\tau^\mu \Big)~, \nn \\
D_\nu^A &=& - 2i\sqrt{m}\p_{\nu}b^A= - 2i\sqrt{m}\, v^A\delta_{\nu}^{\, \tau}~, \\ D^{\mu\dagger}_B &=&  2i\sqrt{m}\p^{\mu}b^\dagger_B= 2i\sqrt{m}\, v^\dagger_B\delta^{\mu}_{\, \tau}~, \nn \\
E^A_B &=& M(\tau)\, \delta^{A}_{B} ~, \nn 
\end{eqnarray}
and
$$M(\tau)\equiv  2m_0 |v|^2\tau^2  + |v|^4\tau^4\,.$$ 
One may note that the last two terms in $C^\mu_\nu$ come from the time-dependent rescaling of gauge fields.  
There is another part in bosons which may be denoted as $\hat{Q}^b_0 + \hat{Q}^b_1$. This has the same form as  $Q^b_0+Q^b_1$ with a replacement $\epsilon^{\mu}_{~ \nu\rho}  \rightarrow -\epsilon^{\mu}_{~ \nu\rho}$. It turns out that they give the identical contributions.

\underline{\it Ghost fields}
\bea 
 Q^g_0 &=& -\Box + m^2_0\,, \qquad \qquad \qquad ~~~ \Box \equiv \p^2_{\tau} + \p^2_{i} \nn \\ &&\nn\\
 Q^g_1(\tau)& =& M(\tau) + \Big(-\frac{ |v|^2}{ 2m_0}+ \frac{5 |v|^4\tau^2}{4m_0^2}\Big)-\frac{ |v|^2 \tau}{m_0}\p_{\tau}\,. \label{ghostD}
\eea
Note that the last two terms in $ Q^g_1(\tau)$ come from the time-dependent rescaling of the ghost fields.

\underline{\it Fermion fields} 

We construct the ``squared'' operator,  $Q^f$, in terms of the original operators by
\[ 
Q^f = \CD\tilde{\CD} = \tilde{\CD}\CD=Q^{f}_{\,0}+ Q^{f}_{\, 1}\,,
\]
where
\bea Q^{f}_{\,0} &=& (-\Box +m^2_0)\delta^{~ \beta}_{\alpha}\delta^{A}_{B}\,,   \nn \\ &&\nn\\
 Q^{f}_{\, 1} &=& M(\tau)\, \delta^{~ \beta}_{\alpha}\delta^{A}_{B} - \slash{\p}^{~ \beta}_{\alpha}(|b|^2\delta^{A}_{B}-2b^{A}b^\dagger_{B})\,. \eea

\section{One-loop corrections}

 The one-loop effective action  is given by %
\[ \Gamma_1 = \sum_{fields} (-1)^\epsilon \ln\det (Q_0 + Q_1) \,, \]
where $\epsilon$ is 0 for the bosonic fields and  1 for the fermion and ghost fields.  
This can be computed systematically using the Schwinger proper time method\footnote{For a review on the Schwinger formalism, for example, see~\cite{Ball:1988xg}.} as
\[ \Gamma_1 =  -\sum (-1)^\epsilon \int^{\infty}_{0} \frac{ds}{s}~\Tr\,   e^{-s(Q_0 + Q_1)}\,, \]
where $\Tr$ denotes sum over all the indices including coordinates.

The one-loop effective potential in the Euclidean space can be read from the one-loop effective action as
\[ \Gamma_1 = \int d^3x V_1(x) \]
and thus  becomes
\[
 V_1(x) = V_1^b +V_1^f + V_1^g = - \sum (-1)^\epsilon\int^{\infty}_{0}\frac{ds}{s}~  \tr\, \langle x | e^{-s(Q_0+Q_1)} |x \rangle \,,
\]
where $\tr$ denotes the trace over gauge group and Lorentz indicies. Standard Dyson perturbative expansion for small $Q_1$ leads to
\bea  \!\!\!\!\!\!\!V_1 &=&   \sum_{n=1}^{\infty} V_{1,n-1}\eea
where
\bea  \!\!\!\!\!\!\!V_{1, n-1} &=&  \sum_{fields}(-1)^\epsilon (-1)^{n} \int^{\infty}_{0}\frac{ds_1\cdots ds_n}{(s_1+\cdots + s_n)}\, \tr\,\langle x| \Big[ e^{-(s_1+\cdots + s_n)Q_0} \prod^{n}_{i=2}Q_1(s_i+\cdots +s_n)\Big] |x \rangle \,,  \nn \eea
where $Q_1(s)$ is defined by
\[
 Q_1(s) \equiv e^{sQ_0}Q_1e^{-sQ_0} \,.
\]

\underline{\it Regularization by dimensional reduction}

The computations typically involve the integration over the momentum $p$ as well as the Schwinger parameters $s_i$. To deal with the divergencies,  we adopt dimensional regularization in the momemtum integrals. It is well known that  Chern-Simons gauge theories  have subtleties in using  dimensional regularization~\cite{Martin:1990xv}\cite{Chen:1992ee}. They arise because the theories are sensitive to the dimension they live in, i.e. Chern-Simons term can be defined only in three dimensions. The same kind of subtleties arises in the general supersymmetric theories as well~\cite{Gates:1983nr} because the number of bosonic/fermionic degrees of freedom are sensitive to the spacetime dimension. 

In order to avoid this kind of problems,  a modified prescription, which is called regularization by dimensional reduction, has been adopted to these theories.   The essential point in this modified version is that the usual dimensional regularization rule, with dimensional continuation from three to $n+1$ dimensions, is applied to the momentum integration with divergencies, while all the contractions in tensor and spinor indices
are performed in three dimensions~\cite{Chen:1992ee}\cite{Gates:1983nr}. For example, we have
\bea \epsilon^{\mu\nu}_{~~ \rho}\epsilon^{\rho}_{~ \lambda\eta} & =& \Big(\delta^{\mu}_{~ \lambda}\delta^{\nu}_{~ \eta} -  \delta^{\mu}_{~ \eta}\delta^{\nu}_{~\lambda}\Big)\,, \nn \\
\delta^{\mu}_{~ \mu} &=& 3\,, 
\eea
in three dimensional Euclidean space.  For the spinor indices, we use
\[ \delta^{\alpha}_{~ \alpha} = 2\,. \]

Now we present the results of our computations. We compute only up to $|v|^4$ terms.
 At first, we consider the gauge/scalar fields contributions to the one-loop effective potential.  In fact it is 
the most nontrivial part in computations. Here we present only the results. For details in calculation, see appendix B.   

\underline{\it Gauge/scalar fields}

As stated earlier, we have two parts of gauge/scalar fields, each from $Q_0^b+Q_1^b$ and  $\hat{Q}_0^b+\hat{Q}_1^b$, which give identical results. Here we present only  $Q_0^b+Q_1^b$ part.  At the end we should double what we have got.
\bea
 V^b_{1,\, 0} &=&  -\Big[\frac{m^2_0}{4\pi}\Big]^{\frac{n+1}{2}}\, 6\Gamma\Big(-\frac{n+1}{2}\Big)\,, 
\nn \\ &&\nn\\
V^b_{1,\, 1}
&=& \frac{1}{m^2_0}\Big[\frac{m^2_0}{4\pi}\Big]^{\frac{n+1}{2}}~ \Gamma\Big(\frac{1-n}{2}\Big)\Big[6M(\tau) + |v|^4\tau^4- \frac{|v|^2}{m_0}+4\frac{|v|^4\tau^2}{m^2_0}\Big ]+ \CO(|v|^6)\,, 
\nn\\&&\nn\\
 V^b_{1,\, 2} 
&=& 
-\frac{1}{m^4_0}\Big[\frac{m^2_0}{4\pi}\Big]^{\frac{n+1}{2}} 
\bigg[ \Gamma\Big(\frac{1-n}{2}\Big)\Big(|v|^4\tau^2 +m^2_0|v|^4\tau^4\Big)
 \nn \\
&& \qquad \qquad \qquad  +\, \Gamma\Big(\frac{3-n}{2}\Big)\Big( 4m_0|v|^2 +  \frac{|v|^4}{2m^2_0} -|v|^4\tau^2+12 m^2_0|v|^4\tau^4\Big) \nn \\
&& \nn \\
&& \qquad \qquad \qquad  +\Gamma\Big(\frac{5-n}{2}\Big)4|v|^4\tau^2  \bigg] + \CO(|v|^6)\,,
\\&&\nn\\
V^b_{1,\, 3} &=& \frac{1}{m^6_0}\Big[\frac{m^2_0}{4\pi}\Big]^{\frac{n+1}{2}}\, \Gamma\Big(\frac{5-n}{2}\Big)\,\Big( 8m^2_0|v|^4\tau^2 -2 |v|^4 \Big)+ \CO(|v|^6)\,,
\nn\\&&\nn\\
V^b_{1,\,4} &=& -\frac{1}{m^8_0}\Big[\frac{m^2_0}{4\pi}\Big]^{\frac{n+1}{2}}\, \Gamma\Big(\frac{7-n}{2}\Big)\, \frac{4}{3}m^2_0|v|^4 + \CO(|v|^6)\,. \nn
\eea

\underline{\it Ghost fields}

There are four identical contributions from ghost fields, $c_{L, R}$, $\tilde{c}_{L, R}$. We denote them  as
\[
 \tr\, {\bf 1}_g = 4 \,.
\]
All together they become 
\bea
V_{1,0}^g
&=& +\Big[\frac{m^2_0}{4\pi}\Big]^{\frac{n+1}{2}}\, \Gamma\Big(-\frac{n+1}{2}\Big)~ \tr\, {\bf 1}_g  \nn\\&&\nn\\
V_{1,1}^g
&=&  -\,  \frac{1}{m^2_0}\Big[\frac{m^2_0}{4\pi}\Big]^{\frac{n+1}{2}} \Gamma\Big(\frac{1-n}{2}\Big) \bigg( M(\tau) - \half \frac{|v|^2}{m_0}+ \frac{5}{4}\frac{|v|^4\tau^2}{m^2_0}\bigg) ~ \tr\, {\bf 1}_g  + \CO(|v|^6)\nn \\&&\nn\\
V_{1,2}^g&=& +\, \frac{1}{m^4_0}\Big[\frac{m^2_0}{4\pi}\Big]^{\frac{n+1}{2}}\Big[\Gamma\Big(\frac{5-n}{2}\Big)  \frac{2}{3}|v|^4\tau^2+  \Gamma\Big(\frac{3-n}{2}\Big)\bigg(2m^2_0|v|^4\tau^4 + \frac{1}{8}\frac{|v|^4}{m^2_0}-|v|^4\tau^2\bigg)  \nn \\
&& \nn \\
&& \qquad \qquad \qquad ~  -\,\frac{1}{4}  \Gamma\Big(\frac{1-n}{2}\Big)\, |v|^4\tau^2 \Big]~ \tr\, {\bf 1}_g + \CO(|v|^6)\,. 
\eea

\underline{\it Fermionic fields}

The computations in fermionic fields are also straightforward. The results are as follows.
\bea
V_{1,0}^f
&=& +\Big[\frac{m^2_0}{4\pi}\Big]^{\frac{n+1}{2}}\, \Gamma\Big(-\frac{n+1}{2}\Big)~ \tr\, {\bf 1}_f\,,
\nn \\&&\nn\\
V_{1,1}^f &=&
   - \frac{1}{m^2_0}\Big[\frac{m^2_0}{4\pi}\Big]^{\frac{n+1}{2}}\,  \Gamma\Big(\frac{1-n}{2}\Big)\,  M(\tau)~ \tr\, {\bf 1}_f \,, \nn \\&&\nn\\
   V_{1,2}^f&=& + \frac{1}{m^4_0}\Big[\frac{m^2_0}{4\pi}\Big]^{\frac{n+1}{2}}\,  \bigg[ \Gamma\Big(\frac{3-n}{2}\Big) \Big(m_0|v|^2  + 2|v|^4\tau^2 + 2m^2_0|v|^4\tau^4\Big) \nn\\ &&\nn\\
&& \qquad \qquad \qquad ~ +  \Gamma\Big(\frac{5-n}{2}\Big)\frac{2}{3} |v|^4\tau^2  \bigg] \tr\, {\bf 1}_f + \CO(|v|^6)\,, \nn \\&&\nn\\
   V_{1,3}^f&=& - \frac{1}{m^6_0}\Big[\frac{m^2_0}{4\pi}\Big]^{\frac{n+1}{2}}\bigg[\Gamma\Big(\frac{5-n}{2} \Big)16m^2_0|v|^4\tau^2\bigg] + \CO(|v|^6)   \,, 
\nn \\&&\nn\\
V_{1,4}^f
&=& + \frac{1}{m^8_0}\Big[\frac{m^2_0}{4\pi}\Big]^{\frac{n+1}{2}}\bigg[\Gamma\Big(\frac{7-n}{2} \Big)\frac{8}{3}m^2_0|v|^4 \bigg]+ \CO(|v|^6)  \,,
\eea
where
\[\tr\, {\bf 1}_f  = \tr\, \delta^{\alpha}_{\, \beta}\delta^{A}_{\, B} =8 \,.\]
%

\underline{\it The results: summary}

By collecting all the results of one-loop effective potentials,  one can easily see that there is a complete cancellation among contributions from the ghosts, fermions and gauge/scalar fields up to order $|v|^2$. 
As a result the one-loop effective potential is given by
\begin{eqnarray*}
 V_{1} & = & V^{b}_1 + V^g_{1} + V^{f}_{1}=    \frac{1}{m^4_0}\Big[\frac{m^2_0}{4\pi}\Big]^{\frac{n+1}{2}}\Gamma\Big(\frac{3-n}{2}\Big) 14 |v|^4\tau^2   \nn \\
&&\qquad \qquad\qquad~~~~- \frac{1}{m^6_0}\Big[\frac{m^2_0}{4\pi}\Big]^{\frac{n+1}{2}}\bigg[\Gamma\Big(\frac{5-n}{2}\Big) 4 +\Gamma\Big(\frac{3-n}{2}\Big)\half \bigg]|v|^4 +\CO(|v|^6)\,. 
\end{eqnarray*}

Our field theory results  correspond to those of a single source brane in the supergravity computations.
The generalization to $N$ source branes is straightforward.   
We simply multiply $N$ factor to $V_1$. 
Now taking $n=2$,  the effective action, including the tree-level part, in field theory side  is finally  given by 
\beq 
\Gamma = \Gamma_{\rm tree} + \Gamma_1 = \int d^3x \bigg(\frac{k}{2\pi}|v|^2 + \frac{7N}{4\pi}\frac{|v|^4\tau^2 }{|b_0|^2}- \frac{5N}{16\pi}\frac{|v|^4}{|b_0|^6} +\CO(|v|^6)\bigg)\,. \eeq
Note that the $|v|^4$ term can be eliminated by a suitable shift of $\tau$ with a time reversal symmetry of the effective action. 

The effective action up to $v^4$ terms obtained from $D=3$ $\CN=8$ super Yang-Mills theory is coincident with the probe action on $AdS_4\times S^7$ in the {\it static gauge}.  In general,  this doesn't have to be the case.

We find  complete agreement between the results from our field theory computations and those from the dual supergravity, if we choose the following gauge for the worldvolume diffeomorphism in the supergravity:
\beq  X^0 = \frac{1}{\sqrt{7}|b_0|^2}\ln ( |b_0|^2 \xi^0)\,, \qquad X^{9} = (\sqrt{7}|b_0|^2 \xi^0)^{-1}\, \xi^1 \,, \qquad X^{10} =\xi^2\,. \eeq
In order to see this,  note that the transverse coordinates $z^A$ are identified with $b^A$ and worldvolume coordinates in supergravity, $(\xi^0, \xi^1,\xi^2)$,  are identified with field theory coordinates, $(t,x^1,x^2)$. Since
 the velocities in supergravity and field theory are defined as
\beq v^{A}_{\rm sugra} \equiv \frac{d z^{A}}{d X^{0}} = \sqrt{7}|b_0|^2\xi^0 \frac{d z^{A}}{d \xi^0} \,, \qquad v^{A}_{\rm field} \equiv \frac{d b^{A}}{d t}\,,  \eeq
one can see that 
\[ v^{A}_{\rm sugra}  =  \sqrt{7} |b_0|^2t~ v^A_{\rm field}\,, \]
and 
\[ S_2 = \int d^3x\Big( \frac{k}{2\pi}|v_{\rm field}|^2 + \frac{7N}{4\pi}\frac{|v_{\rm field}|^4t^2 }{|b_0|^2}\Big)\,. \]
This shows that supergravity results are in complete agreement with  field theory ones.

\section{Conclusion}
In this paper we took a first step toward the understanding of the quantum correction in the ABJM model, which would give a nontrivial test for the $AdS/CFT$ correspondence. We used the $R_\xi$ gauge, which preserves the supersymmetry if the vev is time-independent, to perform one-loop computations. We found complete agreement in membrane scattering dynamics between the results from the ABJM model and those from the dual supergravity on $AdS_4\times S^7/\Z^k$  in a specific gauge for worldvolume diffeomorphism. 

As a result we find that there is no correction in the $v^2$ term. As stated earlier, $\CN=4$ supersymmetry in three dimensions guarantees that the $v^2$ term is one-loop exact. Our result, supplemented with the supersymmetry, shows that there is non-renormalization in the $v^2$ term, i.e. tree-level exact. It would be very nice to show that it is indeed the case by using supersymmetry arguments for $\CN=6$. One may note that this also reflects   the conformal symmetry at the quantum level. 

 We also find that the $v^4$ term appears at one-loop, which agrees with the supergravity computations in the special choice of gauge for worldvolume diffeomorphism. There is a 
non-renormalization theorem, at least for $\CN=8$ supersymmetry, which states that the $v^4$ term appears only at one-loop with possible non-perturbative instanton corrections. Since there is no monopole-instanton for our configurations, we expect our result is exact, at least for $k=1$ and 2. If we start with the background configuration shown in the appendix A, we need to include the instanton corrections to reproduce the results from supergravity. 

It would be very interesting to reexamine the problem using the superfield formalism with the supersymmetric $R_\xi$ gauge.  
It would be also very interesting to see whether there is a, perturbative, non-renormalization theorem for the ABJM model with generic $k$. This might be determined by studying the supersymmetry completion. Another way to see this is to study two-loop corrections to the effective action.

{\bf Acknowledgments}

We would like to thank the theory group at UBC, KEK and KIAS for hospitality. J.H.B, S.H. and S.H.Y were supported by
the Korea Research Foundation Grant funded
by Korea Government(MOEHRD, Basic Reasearch Promotion Fund)
(KRF-2005-070-C00030).
W.J was supported by
 the Science Research Center Program of the
Korea Science and Engineering Foundation through the Center for
Quantum Spacetime \textbf{(CQUeST)} of Sogang University with grant
number R11-2005-021.

\section*{Appendix}

\subsection*{A. Supersymmetry and Gauge Fixing}

In this appendix, we consider different background configurations and the corresponding gauge fixing. 
Consider the  following vacuum expectation values with real $d^A$ and $b^A$
\beq  \bar{Y}^{A} = \left( \ba{cc} d^A & 0  \\  0 & b^A \ea\right)\,, \qquad \bar{Y}^{\dagger}_{A} = \left( \ba{cc} d_{A} & 0  \\  0 & b_A \ea\right)\,. \eeq
The bosonic part in the quadratic Lagrangian is given by
\bea \CL_b = - (y^{\dagger}~ \tilde{y})\left(\ba{cc} \CM & \CN \\ \CQ & \CP \ea\right) {y\choose \tilde{y}^{\dagger}}\,,
\eea
where
\bea \CM^{A}_{~ B}  &=&  \Big[-\Box + (b^2+d^2)^2 -4(b\cdot d)^2\Big]~ \delta^{A}_{\, B} - (b^2+d^2)(b^Ab_B+d^Ad_B) + 2(b\cdot d)(b^Ad_B +d^Ab_B) \,,
\nn \\
\CN^{AB} &=&  (b^2+d^2) (b^Ad^B +d^Ab^B) -2 (b\cdot d)(b^Ab^B+d^Ad^B)\,, \nn \\
\CP_{A}^{~\, B} &=& \CM_{A}^{~\, B}\,, \qquad \CQ_{AB} = \CN_{AB}\,. \eea
The fermionic part is
\bea  - (\psi^{\dagger},\, \tilde{\psi})\, \CD_f \,  { \psi \choose \tilde{\psi}^{\dagger}} \equiv  - (\psi^{\dagger\, A},\, \tilde{\psi}_C)\left(\ba{cc} F^{~ B}_{A} & -2i\epsilon_{ADPQ}b^Pd^Q \\ 2i\epsilon^{CBPQ}b_Pd_Q & - F^{C}_{~ D} \ea\right) { \psi_{B} \choose \tilde{\psi}^{\dagger\, D}} \,, \eea
where 
\[ F^{~ B}_{A}= \Big[-i\slash{\p} + i(b^2-d^2)\Big]~\delta^{~ B}_{A} -2i(b_Ab^{B}-d_{A}d^{B})\,. \] 
For constant $b$ and $d$, one gets
\[ Q_f \equiv \CD_f\CD^{\dagger}_f =\CD_f^{\dagger}\CD_f = \Big[-\Box+(b^2+d^2)^2 -4(b\cdot d)^2\Big]{\bf 1}\,. \]

To get a covariant gauge fixing term which respects the supersymmetry, we introduce
\[  A_{\pm\, \mu} \equiv \half \Big(A_{L\,\mu} \pm A_{R\, \mu}\Big) \,, \]
and take the $R_\xi$ gauge for these gauge fields. The gauge fixing terms are given by
\bea \CL_{GF} &=& -\frac{1}{2\xi_{+}}\Tr\Big(\p_{\mu}A^{\mu}_{+} + \frac{i}{\sqrt{2}}\xi_{+} [\bar{Y}, \delta Y^{\dagger}] +  \frac{i}{\sqrt{2}} \xi_{+}[\bar{Y}^{\dagger}, \delta Y] \Big)^2 \nn \\
&&   -\frac{1}{2\xi_{-}}\Tr\Big(\p_{\mu}A^{\mu}_{-} + \frac{i}{\sqrt{2}}\xi_{-} \{\bar{Y}, \delta Y^{\dagger}\} - \frac{i}{\sqrt{2}} \xi_{-}\{\bar{Y}^{\dagger}, \delta Y\} \Big)^2
\,. \nn \eea
One can show that they are supersymmetric if  $\xi_\pm$ are given by
\[ \xi_{+} = b^2+d^2 + 2\, b\cdot d\,, \qquad \xi_{-} = b^2+d^2 - 2\, b\cdot d\,. \]
Note that $d=0$ case reduces to the same gauge fixing Lagrangian given in the main text.

\subsection*{B. Some details in one-loop computations}

Our normalization conventions for the plane wave basis in the computation of the one-loop effective action are
\[
\langle x | p \rangle  = \frac{1}{(2\pi)^{n+1}}\, e^{ip\cdot x}\,, \] 
with the completeness relations 
\[
\int d^{n+1}x~ |x \rangle  \langle x| = {\bf 1}\,, \qquad    \int d^{n+1}p~ |p \rangle  \langle p| = {\bf 1}\,. \]

Here we present the calculational details of  the bosonic part contributions.   First of all, it is convenient to introduce 
\[ P^{\mu}_{~ \nu} = \delta^{\mu}_{\, \nu} - \frac{p^{\mu}p_{\nu}}{p^2}\,, \qquad R^{\mu}_{~ \nu} = \epsilon^{\mu}_{~ \nu\rho}\frac{p^{\rho}}{p}\,, \]
which give
\beq e^{aP} = 1 - P + e^{a}P\,, \qquad e^{aR} = 1 -P + \cos a\, P + \sin a\, R\,.  \nn \eeq
They have nice properties such as 
\beq P^2 = P\,, \qquad R^2 = -P\,, \qquad PR = RP = R\,, \qquad \tr\, P = 2 \,, \qquad \tr\, R=0\,.  \nn \eeq
Let us  define 
\bea e^{-sQ^{b}_{0}(p)} \equiv \left( \ba{cc} e^{-sA(p)} & 0 \\ 0 & e^{-sB(p)} \ea\right)\,, \eea
where the quadratic operator $A$ and $B$ are given by
\bea A^{\mu}_{\, \nu} &=& p^{\mu}p_{\nu} - m_0 \epsilon^{\mu}_{~ \nu\rho}p^{\rho} + m^2_0\delta^{\mu}_{~ \nu} = (p^2+m^2_0)\delta^{\mu}_{~ \nu} -p^2P^{\mu}_{~ \nu}-m_0pR^{\mu}_{~ \nu}\,, \nn\\
 B &=& (p^2 +m^2_0)\delta^{A}_{B}\,. \nn \eea
Then we obtain
\[ e^{-sA} = e^{-s(p^2+m^2_0)}(1-P) + e^{-sm^2_0}\Big( P\, \cos(smp) + R\, \sin(smp)\Big)\,. \]
The above relations  facilitate the various calculations involving products of $e^{-sQ^b_0}Q^b_1$'s.

The computation of the zeroth order, in $Q_1^b$ insertion,  is straightforward and  goes as follows:
\bea
 V^b_{1,\, 0} &=&  -\int^{\infty}_{0}\frac{ds}{s}\, \tr\, \langle x|  e^{-sQ^b_0}   |x \rangle = -\int^{\infty}_{0}\frac{ds}{s}\int \frac{d^{n+1}p}{(2\pi)^{n+1}}\, \tr\, \langle p|  e^{-sQ^b_0}   |p \rangle \nn\\ 
&=& -\int^{\infty}_{0}\frac{ds}{s}\, e^{-sm^2_0} \int \frac{d^{n+1}p}{(2\pi)^{n+1}}\,  \Big[  5e^{-sp^2} + 2\cos m_0ps\Big] \nn\\
&=&-\Big[\frac{m^2_0}{4\pi}\Big]^{\frac{n+1}{2}}\int^{\infty}_{0} ds  \, e^{-s}\, \bigg[ \frac{5}{s^{(n+3)/2}}   - \frac{4}{s^{n+2}}\frac{\Gamma(n+1)}{\Gamma\big(\frac{n+1}{2}\big)}\sin \frac{n\pi}{2}\bigg]  
\nn \\
&=&
-\Big[\frac{m^2_0}{4\pi}\Big]^{\frac{n+1}{2}}\, 6\Gamma\Big(-\frac{n+1}{2}\Big)\,, 
\eea
where $\tr$ denotes sum over gauge, Lorentz and $SU(4)_R$ indices and we have used 
\[ \frac{\Gamma(n+1)\Gamma(-n-1)}{\Gamma\big(\frac{n+1}{2}\big)} = -\frac{\Gamma\big(-\frac{n+1}{2}\big)}{4\sin\frac{n}{2}\pi }\,. \]
In what follows,  we integrate over Schwinger parameters $s_i$ first, and then calculate momentum integrals. 

The first order part is also straightforward and is given by  
\bea
V^b_{1,\, 1}  &=&  \int^{\infty}_{0}\frac{ds_1ds_2}{s_1+s_2}\, \tr \Big\langle x \Big| e^{-(s_1+s_2)A}C  + e^{-(s_1+s_2)B}E \Big|x \Big\rangle \nn\\
&=&  \int \frac{d^{n+1}p}{(2\pi)^{n+1}}\, \int^{\infty}_{0}\frac{ds_1ds_2}{s_1+s_2}\,  e^{-(s_1+s_2)m^2_0}  \bigg\{\Big[\frac{1}{n+1}\, e^{-(s_1+s_2)p^2} + \frac{n}{n+1}\cos m_0p(s_1+s_2) \Big] \nn \\ && \nn \\
&&\qquad \qquad \qquad ~~~  \times \Big(3M(\tau) - \frac{|v|^2}{m_0}+4\frac{|v|^4\tau^2}{m^2_0}\Big)  + \Big[ p \sin m_0p(s_1+s_2)\Big]2|v|^2\tau^2 + 4M(\tau) \bigg\}      \nn \\
&&\nn \\
&=& \frac{1}{m^2_0}\Big[\frac{m^2_0}{4\pi}\Big]^{\frac{n+1}{2}}~ \Gamma\Big(\frac{1-n}{2}\Big)\Big[6M(\tau) + |v|^4\tau^4- \frac{|v|^2}{m_0}+4\frac{|v|^4\tau^2}{m^2_0}\Big ]\,,
\eea
where we have used 
\[ p_\mu p_\nu = \frac{1}{n+1} p^2 \delta_{\mu\nu}
\]
in the momentum integral.

The calculations of integrals are quite involved  starting from  the second order computations. We present all those integrals in appendix C. 
The second order in the perturbation consists of three parts,
\beq
V^b_{1,\, 2} = [CC]+ [EE]+[DD]\,,
\eeq
where each represents the contribution from the gauge-gauge, the scalar-scalar and the gauge-scalar  fields.
Using those integral formulae given in appendix C, we find the contributions from the second order part as follows: 
\bea 
\big[CC\big] &\equiv& -\int^{\infty}_{0}\frac{ds_1ds_2ds_3}{s_1+s_2+s_3}\, \tr\, \Big\langle x \Big| e^{-(s_1+s_3)A}Ce^{-s_2A}C \Big|x \Big\rangle  \nn \\ 
&=& 
-\frac{1}{m^4_0}\Big[\frac{m^2_0}{4\pi}\Big]^{\frac{n+1}{2}} 
\bigg[ \Gamma\Big(\frac{1-n}{2}\Big)\Big(|v|^4\tau^2 +m^2_0|v|^4\tau^4\Big)
  + \Gamma\Big(\frac{3-n}{2}\Big)\Big(  \frac{|v|^4}{2m^2_0} -5|v|^4\tau^2+ 4|v|^4\tau^4\Big)\nn \\
  &&  \qquad \qquad  \qquad \, \, \, +\Gamma\Big(\frac{5-n}{2}\Big)\frac{4}{3}|v|^4\tau^2  \bigg] + \CO(|v|^6)\,, \nn \\
\big[EE\big] &\equiv& -\int^{\infty}_{0}\frac{ds_1ds_2ds_3}{s_1+s_2+s_3}\, \tr\, \Big\langle x \Big| e^{-(s_1+s_3)B}Ee^{-s_2B}E \Big|x \Big\rangle   \\  
&=&  -\frac{1}{m^4_0}\Big[\frac{m^2_0}{4\pi}\Big]^{\frac{n+1}{2}}\,  \bigg[ \Gamma\Big(\frac{3-n}{2}\Big) 8m^2_0|v|^4\tau^4  +  \Gamma\Big(\frac{5-n}{2}\Big)\frac{8}{3} |v|^4\tau^2  \bigg]+ \CO(|v|^6)\,,  \nn \\ && \nn \\ %
\big[DD\big] &\equiv& -\int^{\infty}_{0}\frac{ds_1ds_2ds_3}{s_1+s_2+s_3}\, \tr\, \Big\langle x \Big| e^{-(s_1+s_3)A}D^{\dagger}e^{-s_2B}D  +  e^{-(s_1+s_3)B}De^{-s_2A}D^{\dagger}  \Big|x \Big\rangle \nn \\ 
 &=& -\frac{1}{m^4_0}\Big[\frac{m^2_0}{4\pi}\Big]^{\frac{n+1}{2}}\, \Gamma\Big(\frac{3-n}{2}\Big) 4(m_0|v|^2 + |v|^4\tau^2) + \CO(|v|^6)\,. \nn \\ && \nn  %
\eea

Similarly, the cubic order can be found to be 
\beq
V^b_{1,\, 3} = [CDD] + [EDD]\,,  \eeq
where
\bea
\big[CDD\big] &\equiv& \int^{\infty}_{0}\frac{ds_1ds_2ds_3ds_4}{s_1+s_2+s_3+s_4}\, \tr\, \Big\langle x \Big| e^{-(s_1+s_4)A}C e^{-s_2A}D^{\dagger} e^{-s_3B}D 
+ e^{-(s_1+s_4)A}D^{\dagger} e^{-s_2B}D e^{-s_3A}C\nn \\ 
&&\qquad \qquad \qquad \qquad \qquad  ~~~ +    e^{-(s_1+s_4)B}De^{-s_2A} Ce^{-s_3A}D^{\dagger}  \Big| x \Big\rangle \nn \\ 
&=& \frac{1}{m^6_0}\Big[\frac{m^2_0}{4\pi}\Big]^{\frac{n+1}{2}}\, \Gamma\Big(\frac{5-n}{2}\Big)\,\Big( 4m^2_0|v|^4\tau^2 -2 |v|^4 \Big)+ \CO(|v|^6)\,, \nn \\ && \nn \\ 
\big[EDD\big] &\equiv&  \int^{\infty}_{0}\frac{ds_1ds_2ds_3ds_4}{s_1+s_2+s_3+s_4}\, \tr\, \Big\langle x \Big|  e^{-(s_1+s_4)B}E e^{-s_2B}D e^{-s_3A}D^{\dagger} +   e^{-(s_1+s_4)B} De^{-s_2A}D^{\dagger} e^{-s_3B} E  \nn \\
&& \qquad \qquad \qquad \qquad \qquad ~~~+  e^{-(s_1+s_4)A}D^{\dagger} e^{-s_2B}E e^{-s_3B}D \Big| x \Big\rangle \nn \\ 
&=& \frac{1}{m^6_0}\Big[\frac{m^2_0}{4\pi}\Big]^{\frac{n+1}{2}}\, \Gamma\Big(\frac{5-n}{2}\Big)\,  4m^2_0|v|^4\tau^2   + \CO(|v|^6)\,. \nn  \eea

Finally, the fourth order one-loop effective potential is given by
\bea 
V^b_{1,\,4} &=& -\int^{\infty}_{0}\frac{\prod_{i=1}^5ds_i}{\sum_{i=1}^5s_i}\, \Big\langle x \Big| e^{-(s_1+s_5)A}D^{\dagger}e^{-s_2B}De^{-s_3A}D^{\dagger}e^{-s_4B}D + (A\leftrightarrow B\,, \, D\leftrightarrow D^{\dagger})\Big|x\Big\rangle \nn \\
&=&-\frac{1}{m^8_0}\Big[\frac{m^2_0}{4\pi}\Big]^{\frac{n+1}{2}}\, \Gamma\Big(\frac{7-n}{2}\Big)\, \frac{4}{3}m^2_0|v|^4 + \CO(|v|^6)\,. \eea
%


%
%
%
%

%
\subsection*{C. Useful Integrals}
In this appendix we collect all the nontrivial integral formulae used. Note that, after the $s_i$ integrations, we are left with the momentum integrals of the form:
\[
 \int \frac{d^{n+1}p}{(2\pi)^{n+1}}\, \frac{p^m}{(p^2+m^2)^r} = \frac{1}{m^{2r-m}}\Big[\frac{m^2}{4\pi}\Big]^{\frac{n+1}{2}}\frac{\Gamma\big(\frac{n+m+1}{2}\big)\Gamma\big(r-\frac{n+m+1}{2}\big)}{\Gamma\big(\frac{n+1}{2}\big)\Gamma(r)}\,.
\]

\noindent Now we present various integral formulae for $s_i$ parameters.

\noindent \underline{\it Symmetric case}
\[ \int^{\infty}_0 \frac{ds_1\cdots ds_n}{(s_1+\cdots + s_n)^m}\, f(s_1+\cdots + s_n)  = \frac{1}{\Gamma(n)} \int^{\infty}_{0} ds\, s^{n-m-1}f(s)\,. \]
\underline{\it Triple integrals over $s_i$ parameters}
\[ \!\!\!\!\!\!\!\!\!\!\!\!\!\!\!\!\!\!\!\!\!\!\!\!\!\!\!\!\!\!\!\!\!\!\!\!\!\!\!\!   \int^{\infty}_{0} \frac{ds_1ds_2ds_3}{s_1+s_2+s_3}\, e^{-(s_1+s_2+s_3)m^2} \Big[\cos mp(s_1+s_3)\,\cos mps_2 \Big] = \half\frac{1}{(p^2+m^2)^2}
\,. \]
\[ \!\!\!\!\!\!\!\!\!\!\!\!\!\!\!\!\!\!\!\!\!\!\!\!\!\!\!\!\!\!\!\!\!\!\!\!\!\!\!\!  \int^{\infty}_{0} \frac{ds_1ds_2ds_3}{s_1+s_2+s_3}\, e^{-(s_1+s_2+s_3)m^2} \Big[\sin mp(s_1+s_3)\,\sin mps_2 \Big] = \half\frac{p^2}{m^2}\frac{1}{(p^2+m^2)^2}
\,. \]
\[\!\!\!\!\! \int^{\infty}_{0} \frac{ds_1ds_2ds_3}{s_1+s_2+s_3}\, e^{-(s_1+s_2+s_3)m^2} \Big[ e^{-s_2p^2}\cos mp(s_1+s_3) + e^{-(s_1+s_3)p^2}\cos mps_2 \Big] = \frac{1}{(p^2+m^2)^2}
\,. \]
\[ \int^{\infty}_{0} \frac{ds_1ds_2ds_3}{s_1+s_2+s_3}\, e^{-(s_1+s_2+s_3)m^2} \Big[ 2\cos mp(s_1+s_2+s_3) + \frac{p}{m} \sin mp(s_1+s_2+s_3) \Big] = \frac{1}{(p^2+m^2)^2}
\,.\]
\bea
&&\!\!\!\!\!\!\!\!\!\!\!\!\!\!\!\!\!\!\!\! \int^{\infty}_{0} \frac{ds_1ds_2ds_3}{s_1+s_2+s_3}\, e^{-(s_1+s_2+s_3)m^2} \bigg[ \Big\{ e^{-(s_1+s_3)p^2}\Big(2\cos mqs_2 + \frac{q}{m}\sin mqs_2\Big) \nn \\
&& \qquad \qquad \qquad\qquad \qquad +\,  e^{-s_2q^2}\Big(2\cos mp(s_1+s_3) + \frac{p}{m} \sin mp(s_1+s_3)\Big) \Big\} + \Big\{ p \leftrightarrow q \Big\}  \bigg]  \nn \\
&& \qquad \qquad \qquad ~~~~~  =\,  \frac{2}{(p^2+m^2)(q^2+m^2)} + \frac{1}{m^2}\Big[\frac{1}{p^2+m^2} + \frac{1}{q^2+m^2}\Big]
\,. \nn
\eea
\[ \int^{\infty}_{0} \frac{ds_1ds_2ds_3}{s_1+s_2+s_3}\, e^{-(s_1+s_2+s_3)m^2} \Big[ e^{-(s_1+s_3)p^2 -s_2q^2}  +  e^{-(s_1+s_3)q^2 -s_2p^2}  \Big]   =  \frac{1}{(p^2+m^2)(q^2+m^2)}
\,. \]
\bea 
&&\!\!\!\!\!\!\!\!\!\!\!\!\!\!\!\!\!\!\!\!\!\!\!\!\!\!\!\!\! \int^{\infty}_{0} \frac{ds_1ds_2ds_3}{s_1+s_2+s_3}\, e^{-(s_1+s_2+s_3)m^2} \bigg[ \Big\{ \Big(2\cos mp(s_1+s_3) + \frac{p}{ m}\sin mp(s_1+s_3)\Big) \nn \\
&& \qquad \qquad \qquad \qquad \qquad \qquad   \times \, \Big(2\cos mqs_2 + \frac{q}{m} \sin mqs_2 \Big) \Big\}  + \Big\{ p \leftrightarrow q \Big\}  \bigg]    \nn \\
&& \qquad \qquad \qquad ~~~ =\,  \frac{p^2q^2}{m^4}\frac{1}{(p^2+m^2)(q^2+m^2)} + \frac{2}{m^2}\Big[\frac{1}{p^2+m^2} + \frac{1}{q^2+m^2}\Big]
\,. \nn
\eea
\bea 
&&\!\!\!\!\!\!\!\!\!\!\!\!\!\!\!\!\!\!\!\!\!\!\!\!\!\!\!\!\!\!\!\! \int^{\infty}_{0} \frac{ds_1ds_2ds_3}{s_1+s_2+s_3}\, e^{-(s_1+s_2+s_3)m^2} \bigg[ \Big\{ \Big(2\sin mp(s_1+s_3) - \frac{p}{m}\cos mp(s_1+s_3)\Big) \nn \\
&& \qquad \qquad \qquad \qquad \qquad \qquad   \times \, \Big(2\sin mqs_2 - \frac{q}{m} \cos mqs_2 \Big) \Big\}  + \Big\{ p \leftrightarrow q \Big\}  \bigg]    \nn \\
&& \qquad \qquad \qquad ~~~ =\,  \frac{pq}{m^2}\frac{1}{(p^2+m^2)(q^2+m^2)}
\,. \nn
\eea

\underline{\it Quadruple integrals over $s_i$ parameters}
\bea  &&\int^{\infty}_{0}\frac{\prod_{i=1}^{4}ds_i}{\sum_{i=1}^{4}s_i}e^{-m^2\sum_{i=1}^{4}s_i}\bigg\{   e^{-(s_2+s_3)p^2}\cos mp(s_1+s_4) +  e^{-(s_1+s_3+s_4)p^2}\cos mps_2 \nn \\
&& \qquad \qquad \qquad \qquad \qquad \qquad \qquad \qquad  ~~~ +\, e^{-(s_1+s_2+s_4)p^2} \cos mps_3  \bigg\}  = \frac{1}{(p^2+m^2)^3}\,.  \nn 
\eea
\bea  &&\int^{\infty}_{0}\frac{\prod_{i=1}^{4}ds_i}{\sum_{i=1}^{4}s_i}e^{-m^2\sum_{i=1}^{4}s_i}\bigg\{   e^{-s_2 p^2}\Big[\cos mp(s_1+s_3+s_4) + \frac{p}{2m}\sin mp(s_1+s_3+s_4)\Big]  \nn \\
&& \qquad \qquad \qquad ~~~~~  +\,  e^{-s_3p^2}\Big[\cos mp(s_1+s_2+s_4) + \frac{p}{2m}\sin mp(s_1+s_2+s_4)\Big] \nn \\
&& \qquad \qquad \qquad ~~~~~  +\, e^{-(s_1+s_4)p^2}\Big[ \cos mp(s_2+s_3) + \frac{p}{2m}\sin mp(s_2+s_3)\Big] \bigg\}  = \frac{1}{(p^2+m^2)^3}\,.  \nn 
\eea
\underline{\it Quintic integrals over $s_i$ parameters}
\bea  &&\int^{\infty}_{0}\frac{\prod_{i=1}^{5}ds_i}{\sum_{i=1}^{5}s_i}e^{-m^2_0\sum_{i=1}^{5}s_i}\bigg\{ \Big[ e^{-(s_1+s_5)p^2}\cos mps_3 + e^{-s_3p^2}\cos mp(s_1+s_5)\Big]e^{-(s_2+s_4)p^2} \nn \\
&& \qquad \qquad \qquad \qquad \qquad  +\, \Big[ e^{-s_2p^2}\cos mps_4 + e^{-s_4p^2}\cos mps_2 \Big]e^{-(s_1+s_3+s_5)p^2} \bigg\}  = \frac{1}{(p^2+m^2)^4}\,.  \nn 
\eea
\bea  &&\int^{\infty}_{0}\frac{\prod_{i=1}^{5}ds_i}{\sum_{i=1}^{5}s_i}e^{-m^2_0\sum_{i=1}^{5}s_i}\bigg\{ \Big[  \cos mps_3 \cdot \cos mp(s_1+s_5)\Big]e^{-(s_2+s_4)p^2} \nn \\
&& \qquad \qquad \qquad \qquad \qquad   +\, \Big[ \cos mps_2 \cdot \cos mps_4 \Big]e^{-(s_1+s_3+s_5)p^2} \bigg\}  = \half \frac{1}{(p^2+m^2)^4}\,.  \nn 
\eea


\end{document}